\documentclass[aps,prd,twocolumn,nofootinbib,preprintnumbers]{revtex4} 
\usepackage[dvipdfmx]{graphicx}
\usepackage{bm,latexsym,amsmath,amssymb,amsfonts,mathrsfs}
\usepackage[normalem]{ulem} 
\usepackage{cancel}
\usepackage[usenames]{color}

\newcommand{\del}{\partial}

 \renewcommand\sout{\bgroup \color{red} \ULdepth=-.5ex \ULset}

\begin{document}
\title{Vacua by Derivative Corrections in 
$\mathcal{N} = 1$ Supergravity with Matter Multiplets 
}

\author{Atsuki Inoue}
\email[Email: ]{ms21802"at"st.kitasato-u.ac.jp}

\author{Shin Sasaki}
\email[Email: ]{shin-s"at"kitasato-u.ac.jp}

\affiliation{%
Department of Physics, Kitasato University, Sagamihara 228-8555, Japan}

\begin{abstract}
We study the vacuum structures of four-dimensional $\mathcal{N} = 1$
old minimal supergravity with higher derivative corrections.
We find that $\mathcal{N} = 1$ supergravity with Riemann
 curvature square corrections and higher derivative matter chiral multiplets
 induces a non-trivial de Sitter vacuum, even in the absence of
 superpotentials.
This vacuum generically breaks supersymmetry.
We show that the auxiliary fields in the gravity and the chiral multiples play
 important roles to generate a potential in supersymmetric higher
 derivative theories.
\end{abstract}

\maketitle

\section{Introduction} \label{sect:introduction}

Vacuum structures in supergravity theories have been intensively
studied due to their phenomenological importance.
They are highly constrained because of the restricted scalar potential properties.
For example, four-dimensional $\mathcal{N} = 1$ supergravities exhibit
scalar potentials of the generic form $V = e^{K} ( g^{ij^*} D_i W D_{j^*}
\bar{W} - 3 |W|^2 )$ possibly with D-term contributions. 
Here $K, g_{ij}$ are the K\"ahler potential and the K\"ahler
metric, $D_i$ is the covariant derivative in the K\"ahler geometry and
$W$ is a superpotential. 
A widely known fact that any de Sitter vacua in $\mathcal{N} = 1$ supergravity break supersymmetry is based on
this potential structure.
In order to admit (meta)stable de Sitter vacua, which are supposed to
describe our universe, appropriate K\"ahler and
superpotentials have to be prepared.
However, this is not always possible when supergravities are viewed as low-energy
effective theories of UV theories, like string theories \cite{Gukov:1999ya, Kachru:2003aw,
Kachru:2003sx, Denef:2004ze, Denef:2004cf}.

On the other hand, low-energy effective theories generically receive
derivative corrections.
For example, the fourth order derivative corrections to the gravity sector
are given by the curvature square terms.
It is known that the bosonic $\mathcal{R} + \mathcal{R}^2$ gravity contains a real propagating degree of freedom known as
the scalaron \cite{Whitt:1984pd}. Here $\mathcal{R}$ is the Ricci scalar.
Due to the new scalar degree of freedom appearing in the derivative
corrections, there are room for the modifications of the scalar
potentials. Indeed, the bosonic $\mathcal{R} + \mathcal{R}^2$ gravity is equivalently dual
to the Einstein gravity with a real scalar field accompanied by a non-trivial
potential. 
This was used to realize an inflation model \cite{Starobinsky:1980te}.
In the supersymmetric counterpart, this was first discussed in
\cite{Ferrara:1978rk} and supersymmetric completions of the $\mathcal{R} + \mathcal{R}^2$
gravity in the old minimal \cite{Cecotti:1987sa} and the new minimal
frameworks \cite{Cecotti:1987qe} have been studied.
For the former, the scalar and the vector auxiliary fields $M$,
$\nabla^m b_m$ become propagating. In the dual picture they together
with the scalaron form two chiral multiplets $S$ and $T$ while the dual
for the latter contains a vector multiplet.
The difference of the auxiliary fields in $\mathcal{N} = 1$ supergravity
traces back to the gauge fixing in the superconformal tensor calculus
\cite{Kugo:1982cu}.
 They are a class of $f(\mathcal{R}, \bar{\mathcal{R}})$ supergravities,
 where $f$ is an arbitrary function of the curvature superfields
 $\mathcal{R}, \bar{\mathcal{R}}$.
 Vacuum structures of $f(\mathcal{R},\bar{\mathcal{R}})$ pure supergravity
 \cite{Hindawi:1995qa} and with the chiral matter multiplets
 \cite{Hindawi:1996qi} have been studied.

Things get more involved when higher derivative corrections in chiral
matter sectors are introduced.
It is known that supersymmetric completions of higher derivative matter
terms, even in the absence of the gravity sector, are generically cumbersome issue.
This is mainly due to the auxiliary field problem \cite{Gates:1995fx,
Gates:2000rp} and the presence of the Ostrogradski's ghosts \cite{Ostrogradski:1850}.
The former stems from the fact that the equation of motion for the
auxiliary field $F$ in the chiral multiplet ceases to be algebraic.
The latter is a fate of higher spacetime derivatives of fields of the
form $\del^n \varphi$ ($n\ge 2$).
The higher derivative chiral model \cite{Khoury:2010gb, Khoury:2011da, Koehn:2012te} is a supersymmetric completion of higher derivative
scalar models that does not suffer from these problems.
This is given by the form $U(\Phi, \bar{\Phi}) D_{\alpha} \Phi D^{\alpha} \Phi
\bar{D}^{\dot{\alpha}} \Phi^{\dagger} \bar{D}_{\dot{\alpha}}
\Phi^{\dagger}$ in the D-term.
Here $D,\bar{D}$ are the $\mathcal{N} = 1$ supercovariant derivatives
and $\Phi$ is a chiral superfield, $U(\Phi, \bar{\Phi})$ is an
arbitrary function of $\Phi, \bar{\Phi}$ and the spacetime derivatives of them.
This has been utilized to write down a supersymmetric Dirac-Born-Infeld model
\cite{Rocek:1997hi, Sasaki:2012ka}, Skyrme models \cite{Adam:2011hj, Adam:2013awa, Gudnason:2015ryh,
Gudnason:2016iex} and to study nonlinear realizations \cite{Nitta:2014fca}, BPS
states \cite{Nitta:2014pwa, Nitta:2015uba, Nitta:2020gam} and so on.

It is remarkable that the higher derivative chiral model admits
non-trivial scalar potentials even in the absence of superpotentials.
A rich structure of scalar potentials arises from the non-trivial
solutions to the equation of motion for the auxiliary field.
As we will show in below, this mechanism is in contrast to the situation in the gravity
sector in which explicit scalar potentials appear in the dual frame.
The higher derivative chiral model coupled with the Einstein supergravity has been
studied \cite{Koehn:2012ar, Farakos:2012qu}.
It is shown that the scalar potential induced by the higher derivative
chiral model is negative semi-definite which results in anti-de Sitter or Minkowski vacua.
They are uplifted by the D-term contributions from the gauge sector.

In this letter, we study vacuum structures of a four-dimensional
$\mathcal{N} = 1$ model including the fourth order derivative
corrections both in the gravity and the chiral matter sectors.
We show that supergravity with curvature square terms coupled with the higher derivative chiral
model induces non-trivial scalar potential and exhibits a de Sitter vacuum
even in the absence of superpotentials and gauge sectors.
This is in contrast to the models in \cite{Koehn:2012ar, Farakos:2012qu}
and shows that derivative corrections to the gravity sector can uplift
the anti-de Sitter vacua in a natural way.

\section{Forth order derivatives in $\mathcal{N} = 1$
 supergravity with chiral matter} \label{sect:hd_chiral_model}
We consider the fourth order derivative corrections to four-dimensional
$\mathcal{N} = 1$ supergravity coupled with the chiral matter
multiplets.
In the following, we follow the Wess-Bagger convention \cite{Wess:1992cp}.
The off-shell gravity and the chiral multiplets contain bosonic fields $(e_m^a,
M, b_m)$ and $(A, F)$ respectively.
Here $e_m^a, A,$ are the vierbein and the complex scalar field, 
$M,b_m,F$ are the auxiliary fields.
In the fourth derivative order, the Lagrangian of the gravity sector is
given in the form of curvature square terms.
Supersymmetric completions of the curvature square terms have been
studied \cite{Theisen:1985jr}. For example, we have the $\mathcal{R}^2$ term as 
\begin{align}
\mathcal{L} =& \ \int \! d^2 \Theta \, 2 \mathcal{E}
 (\bar{\mathcal{D}}^2 - 8 R) R \bar{R}
\notag \\
=& \ - \frac{1}{18} e 
\Big[
\mathcal{R}^2 
- 4 (\del_m M) (\del^m \bar{M}) 
+ 4 (\nabla_m b^m)^2 
\notag \\
& \ 
- \frac{4}{3} i b^m (\bar{M} \del_m M - M \del_m \bar{M}) 
+ \frac{4}{9} (M \bar{M})^2 
\notag \\
& \ 
+ \frac{4}{9} (b_m b^m)^2 + \frac{4}{9} M
 \bar{M} b_m b^m - \frac{4}{3} \mathcal{R} b_m b^m - \frac{2}{3}
 \mathcal{R} M \bar{M}
\Big],
\end{align}
where $2 \mathcal{E}$ is a chiral density superfield
involving the determinant of the vierbein $e = \det e_m {}^a$, 
$\mathcal{D}, \bar{\mathcal{D}}$ are supercovariant derivatives in
the curved space and $R$ is the curvature superfield.
We have shown only the bosonic component fields and 
employ the convention for the gravitational constant $\kappa = 1$ in the following.
As is well-known, the auxiliary fields $M,b_m$ get the kinetic term and
they become propagating. Therefore they are not integrated out but
should be treated independent dynamical fields.
Indeed, the curvature square theory is equivalent to 
the Einstein gravity coupled with scalar fields.
Although it is possible to rewrite the Lagrangian via a duality
\cite{Cecotti:1987sa, Cecotti:1987qe}, 
we never switch to the dual form since we keep the derivative
corrections be manifest.

For the matter sector, we consider the higher derivative
chiral model. 
For simplicity, we focus on a model with a single chiral superfield.
The general curvature square invariants involve the Ricci and Riemann tensors.
Then the superfield matter Lagrangian coupled with the gravity sector is defined as

\begin{align}
  \mathcal{L} = 
  &\frac{3}{8} \int \! \! d^2 \Theta \, 2 \mathcal{E}
  \left(
  \bar{\mathcal{D}} \bar{\mathcal{D}} - 8 R
  \right)
  \!
  \Bigg[
  e^{- \frac{K}{3}}
  + \alpha
  \left(
  R \bar{R} - \frac{1}{4} G_{\alpha \dot{\alpha}} G^{\alpha \dot{\alpha}}
  \right)
  \notag \\
  &
  - \frac{1}{3} U (\Phi, \bar{\Phi})
  \mathcal{D}^{\alpha} \Phi \mathcal{D}_{\alpha} \Phi
   \bar{\mathcal{D}}^{\dot{\alpha}} \Phi^{\dagger}
  \bar{\mathcal{D}}_{\dot{\alpha}}
   \Phi^{\dagger}
  \Bigg] 
  \notag \\
  &+\int \! \! d^2 \Theta \, 2 \mathcal{E}(- \gamma W_{\alpha \beta \gamma} W^{\alpha \beta \gamma})
  + \mathrm{h.c.},
  \label{eq:superfield_Lagrangian}
  \end{align}
where the curvature superfields $G_{\alpha \dot{\alpha}} G^{\alpha \dot{\alpha}}$ and $W_{\alpha \beta
\gamma} W^{\alpha \beta \gamma}$ contain the Ricci and the Weyl tensor squares, respectively
\cite{Theisen:1985jr}, and $\alpha, \gamma$ are free parameters.
We again stress that the Lagrangian \eqref{eq:superfield_Lagrangian} does not
contain superpotentials.
After the Weyl rescaling, we have the component Lagrangian for the bosonic fields;
\begin{align}
e^{-1} \mathcal{L} =& \ 
- \frac{1}{2} \mathcal{R} 
- \frac{3}{4} e^{\frac{2K}{3}} \del_m
 (e^{-\frac{K}{3}}) \del^m (e^{-\frac{K}{3}})
\notag \\
& \ 
- 
\Big( 
g_{A \bar{A}} - \frac{1}{3} K_A K_{\bar{A}}
\Big) g^{mn} \del_m A \del_n \bar{A}
+
g_{A \bar{A}} e^{\frac{K}{3}} F \bar{F}
\notag \\
& \ 
-
\frac{1}{3} e^{\frac{K}{3}}
\Big(
M \bar{M} + M F K_A + \bar{M} \bar{F} K_{\bar{A}} + F \bar{F} K_A K_{\bar{A}}
\Big)
\notag \\
& \ 
-
\frac{\alpha}{16} 
\Big(
\mathcal{R}^2 - 3 \mathcal{R}_{mn} \mathcal{R}^{mn}
\Big)
-
\frac{\gamma}{8} \mathcal{C}_{mnkl} \mathcal{C}^{mnkl}
\notag \\
& \ 
+
16 U (A,\bar{A}) 
\Big[
(\del_m A \del^m A)(\del_n \bar{A} \del^n \bar{A})
\notag \\
& \ 
- 2 e^{\frac{K}{3}} F \bar{F} (\del_m A \del^m \bar{A})
+ e^{\frac{2K}{3}} F^2 \bar{F}^2
\Big] + \cdots,
\label{eq:component_Lagrangian}
\end{align}
where $g_{A \bar{A}} = \frac{\del^2 K}{\del A \del \bar{A}}$, 
$K_A = \frac{\del K}{\del A}, K_{\bar{A}} = \frac{\del
K}{\del \bar{A}}$, and $\mathcal{R}_{mn}, \mathcal{C}_{mnkl}$ are the Ricci and the
Weyl tensors. The Weyl tensor contains $\mathcal{R}$, $\mathcal{R}_{mn}$
and the Riemann tensor $\mathcal{R}_{mnkl}$.
The last dots are the derivative terms of $M, b_m$ and a total
derivative term, which are irrelevant in our discussions.
Note that the relative coefficient between $R \bar{R}$ and $G_{\alpha \dot{\alpha}} G^{\alpha \dot{\alpha}}$
in \eqref{eq:superfield_Lagrangian}
has been chosen so that the terms like $ M \bar{M} \mathcal{R}$ are absent in
the Lagrangian and the Einstein frame is retained.
Although higher curvature corrections other than the $f
(\mathcal{R})$-type may involve the Ostrogradski's ghosts 
accompanying the negative kinetic terms of the graviton \cite{Stelle:1977ry},
we never care about these in the following.
This is because the negative kinetic energy of 
graviton is highly suppressed by the Planck mass $M_{\mathrm{p}}$ and
the ghost plays no role in determining the low-energy geometry in the vacuum.

Particular emphasis is placed on the auxiliary field $F$ in the higher
derivative chiral model. The equation of motion for $\bar{F}$ is given
by 
\begin{align}
 g_{A \bar{A}} e^{\frac{K}{3}} F - \frac{1}{3} e^{\frac{K}{3}}
 K_{\bar{A}} (\bar{M} + F K_A) + 32 U e^{\frac{2K}{3}} F F \bar{F} =
 0.
\label{eq:eom_F}
\end{align}
Obviously, this is no longer the linear equation but the cubic order in $F$.
Despite this fact, this is an algebraic equation and solutions are obtained by
the Cardano's formula \cite{Sasaki:2012ka}.
Since we are interested in the vacuum structures of the model, we assume
that $\del_m A = b_m = \del_m M = 0$ and the vanishing fermions in the following.
Then the solutions to the equation \eqref{eq:eom_F} are given by
\begin{align}
F \bar{F} =& \ 
\omega^a \sqrt[3]{- \frac{q}{2} + \sqrt{ \left(\frac{q}{2} \right)^2 +
 \left( \frac{p}{3} \right)^3}} 
\notag \\
& \ 
+ \omega^{3-a} \sqrt[3]{- \frac{q}{2} - \sqrt{\left(\frac{q}{2}\right)^2
 + \left( \frac{p}{3}\right)^3}}
+ \frac{2}{3} \frac{e^{- \frac{K}{3}}}{32 U} G_{A \bar{A}},
\label{eq:FF_sol}
\end{align}
where $\omega^3 = 1$, $a=0,1,2$ and 
we have defined
\begin{align}
G_{A \bar{A}} =& \ g_{A \bar{A}} - \frac{1}{3} K_A K_{\bar{A}},
\notag \\
p =& \ - \frac{1}{3} \left( \frac{e^{-\frac{K}{3}}}{32 U} \right)^2
 G^2_{A \bar{A}},
\notag \\
q =& \ - \frac{1}{9} \left( \frac{e^{-\frac{K}{3}}}{32 U} \right)^2
 K_A K_{\bar{A}} M \bar{M} - \frac{2}{27} \left( \frac{e^{-
 \frac{K}{3}}}{32 U} \right)^3 G^3_{A \bar{A}}.
\end{align}
We note that the formal expression \eqref{eq:FF_sol} is justified only
when $F \bar{F} \ge 0$, otherwise there are no solutions.
The fact that there are generically three independent solutions \eqref{eq:FF_sol} for $a=0,1,2$
causes several consequences in the on-shell Lagrangian.
First, there are three distinct on-shell branches 
associated with the three solutions $a=0,1,2$ \cite{Nitta:2014pwa,
Nitta:2015uba, Nitta:2020gam}.
Second, the fourth order term of $F$ in the Lagrangian induces
non-trivial scalar potential even in the absence of superpotentials $W$.
Indeed, we find that the scalar potential is given by
\begin{align}
V (x,y) =& \  \frac{1}{3} e^{\frac{K}{3}} x 
+ e^{\frac{K}{3}} 
\Big(
g_{A \bar{A}}- \frac{1}{3} K_A K_{\bar{A}}
\Big) F \bar{F} (x,y)
\notag \\
& \ 
+
48 U e^{\frac{2 K}{3}} (F \bar{F} (x,y))^2.
\label{eq:scalar_potential}
\end{align}
Here $x = M \bar{M}, y = A \bar{A}$ and $F\bar{F}(x,y)$ is a solution in
\eqref{eq:FF_sol}.
Although the scalar potential \eqref{eq:scalar_potential} is independent
of $\alpha$ and $\gamma$, it vanishes when the fourth derivative
interactions are absent.
Namely, when $\alpha = \gamma = U = 0$, we have $F = M = 0$ and $V$ becomes trivial.
For $\alpha = \gamma = 0$ but $U \not= 0$ case, we have a non-trivial $F$ but the scalar
field $M$ remains to be auxiliary and it is integrated out. 
The resulting scalar potential is a negative semi-definite and allows only for
anti-de Sitter (or Minkowski) vacua \cite{Farakos:2012qu}.

\section{Vacuum Structures} \label{sect:hd_sugra}

\subsection{Potential minima}
We now examine a minimum of the scalar potential \eqref{eq:scalar_potential}.
In the following, we consider the canonical K\"ahler potential of the
simplest form $K = A\bar{A}$ and $U = \text{const.} = \beta$. Even in such a case, the scalar potential
still possesses non-trivial structure.
Since $M$ is not an auxiliary field anymore, vacua are defined by minima both in the $x = M \bar{M}$
and $y = A \bar{A}$ directions.
We also note that the solutions \eqref{eq:FF_sol} are allowed when $F
\bar{F}$ is positive semi-definite.
This means that the range of VEVs of the scalar fields $(x,y)$ is
restricted.
Since the functional form $F \bar{F} (x,y)$ is different for
$a= 0,1,2$ branches, we study each branch separately.

\subsubsection{$a=0$ branch}
First we focus on the $a=0$ branch. 
This is always a real solution.
For a fixed $x$ and large values of $y$, the dominant contribution comes from the
first term in \eqref{eq:scalar_potential} and the potential grows to infinity:
\begin{align}
V (x,y) \sim \left(\frac{1}{\beta^2}\right)^{\frac{1}{3}}
 e^{\frac{y}{9}} (y x)^{\frac{1}{3}} \to \infty,
\quad (x\not=0, y \to \infty).
\end{align}
The same behaviour holds even for a fixed $y$ and large values of $x$.
For a fixed $x$ and $y \sim 0$, we find
\begin{align}
V (x,y) \sim \frac{1}{24 \beta} (3 + 8 \beta x) + \frac{1}{12 \beta}
 (1-8 \beta x) y,
\quad (y \to 0).
\end{align}
Thus for $\beta > 0$, $5 - 40x <0$ or for $\beta < 0$, $5-40x >0$,
the potential decreases along the
$y$-direction and then it increases again as we go to $y \to \infty$.
Then we expect that there are minima in the $a=0$ branch.
A numerical analysis helps us to find a minimum of the potential.

\begin{figure}[t]
\begin{center}
\includegraphics[scale=.5]{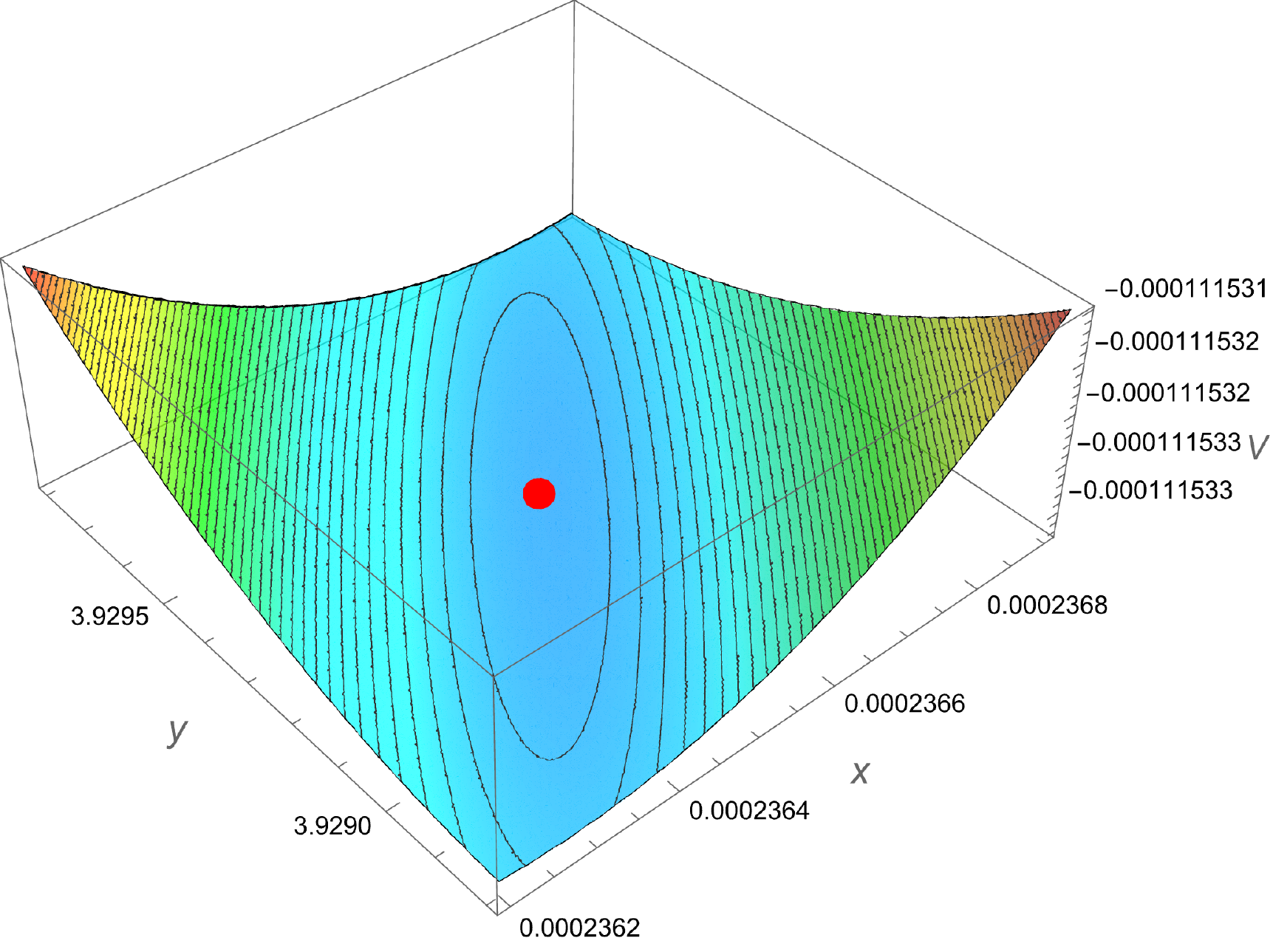}
\end{center}
\caption{The scalar potential in the $a=0$ branch. 
$\beta = 1$.
The red point indicates a minimum.}
\label{fig:a0potential_positive}
\end{figure}

\begin{figure}[htbp]
\begin{center}
\includegraphics[scale=.4]{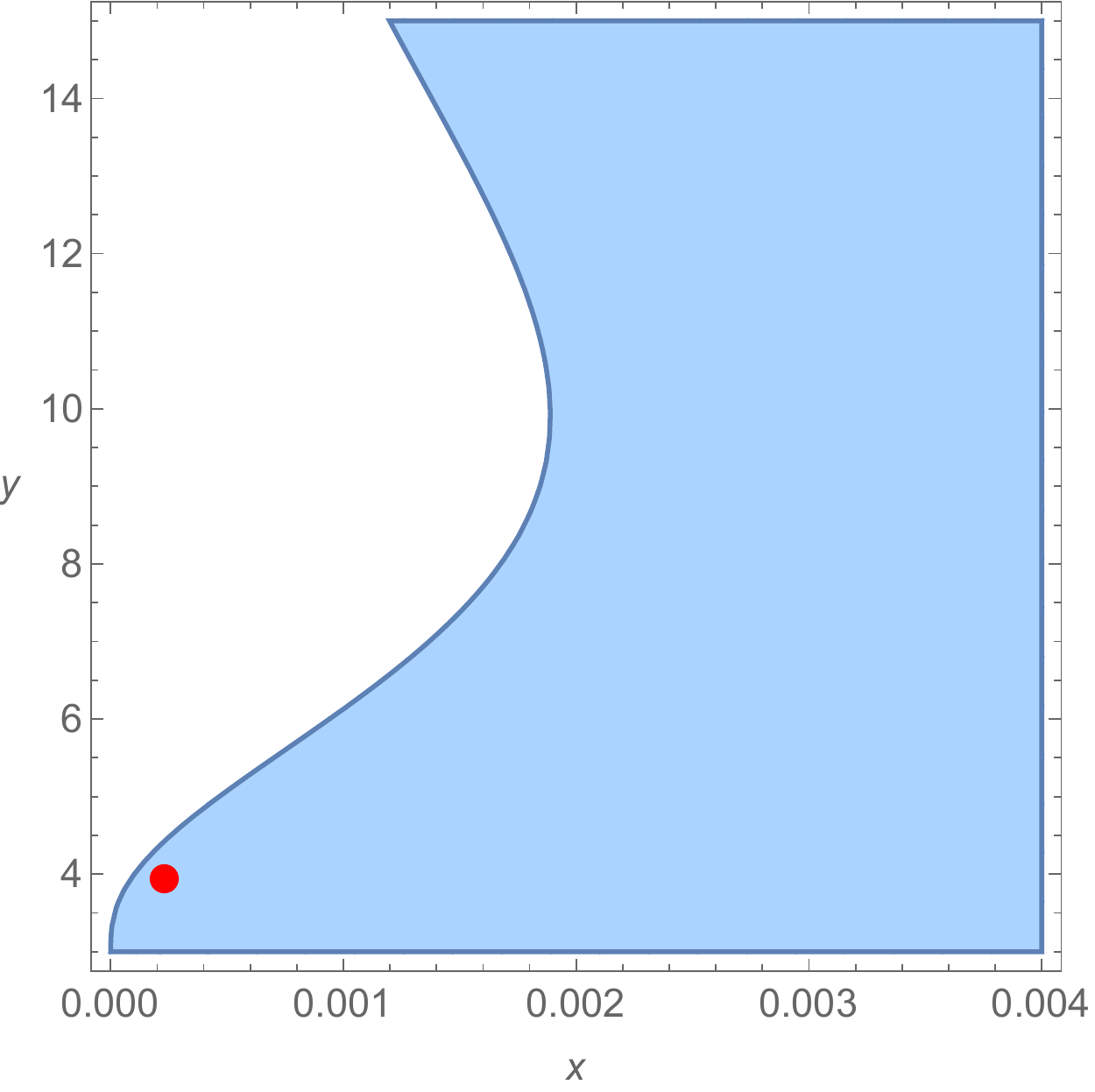}
\end{center}
\caption{The allowed region of the VEV $(x,y)$ for $F \bar{F} (x,y) \ge 0$.
The red point represents the minimum of the potential (see Fig \ref{fig:a0potential_positive}).}
\label{fig:FFpositive_a0}
\end{figure}

For $\beta > 0$ case, the result is given in Fig \ref{fig:a0potential_positive}.
We find a minimum where the vacuum energy is negative, $\Lambda = -
1.1153 \times 10^{-4}$.
The eigenvalues of the mass matrix
\begin{align}
M^2 = 
\frac{1}{2}
\left(
\begin{array}{cc}
\frac{\del^2 V}{\del \varphi_i \del \bar{\varphi}_j} & 
\frac{\del^2 V}{\del \varphi_i \del \varphi_j} \\
\frac{\del^2 V}{\del \bar{\varphi}_i \del \bar{\varphi}_j} &
\frac{\del^2 V}{\del \bar{\varphi}_i \del \varphi_j}
\end{array}
\right), \quad \varphi_i = (A,M),
\end{align}
at the vacuum are given by 
$m^2 = (5.2979, 4.3826 \times 10^{-3}, 0, 0)$
which guarantees that there are no
tachyonic modes. The two zero-eigenvalues correspond to the Nambu-Goldstone modes
associated with the phase directions of $M$ and $A$.
Since the supersymmetry transformation of the chiral fermion is
proportional to the auxiliary field $F$ and  
$F \bar{F}| = 4.8905 \times 10^{-4} \not= 0$, we find that the vacuum breaks supersymmetry.
We also note that the minimum is in the region where the solution $F \bar{F} \ge 0$
is justified. See FIG. \ref{fig:FFpositive_a0} for the allowed
region of VEVs $(x,y)$.

\begin{figure}[t]
\begin{center}
\includegraphics[scale=.45]{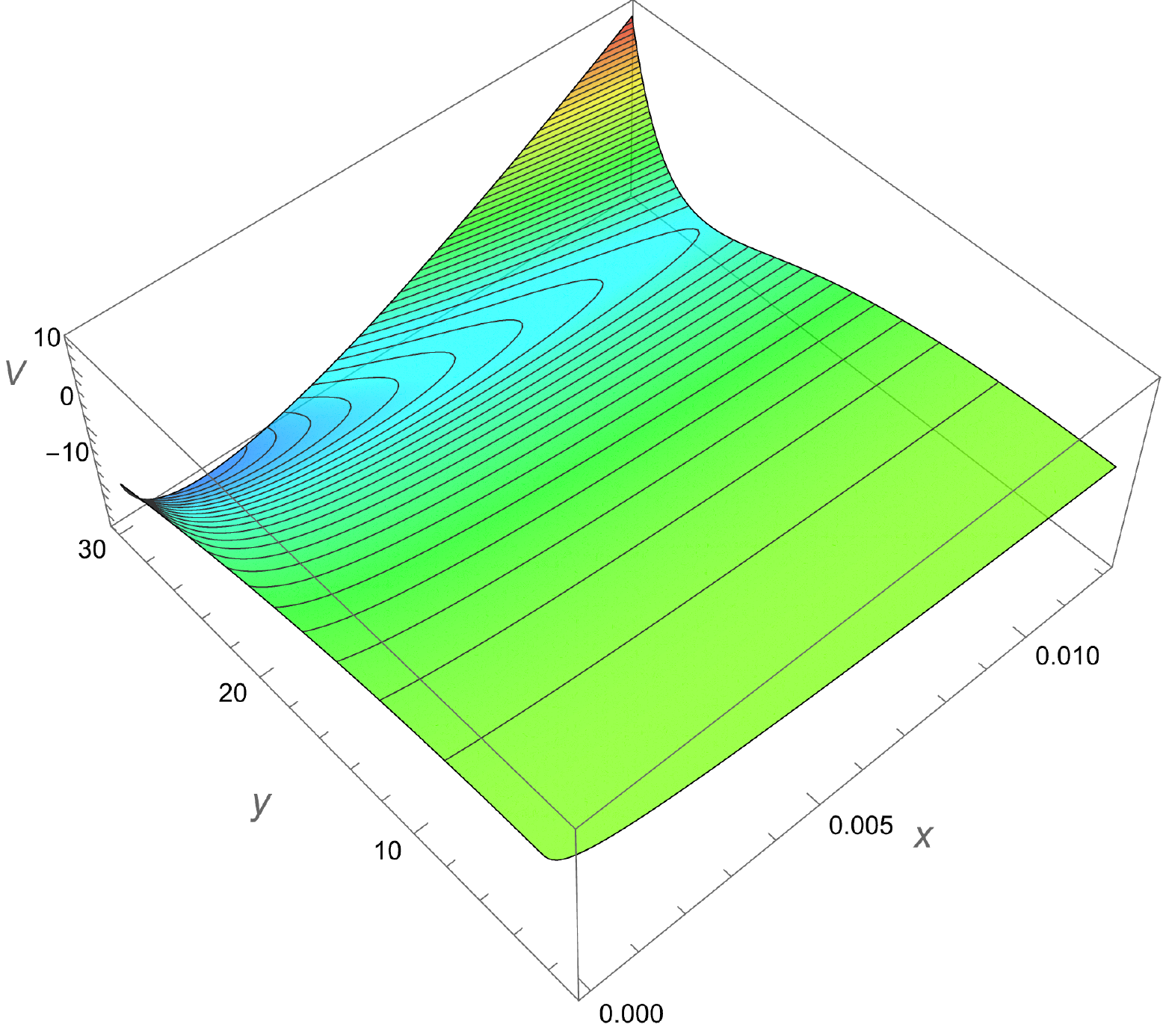}
\end{center}
\caption{The scalar potential in the allowed region of $F \bar{F} \ge 0$.
The $a=0$ branch for $\beta = -1$.
}
\label{fig:a0potential_negative}
\end{figure}

For $\beta < 0$, however, we find that the scalar potential is unbounded
from below. It becomes infinitely small along the $x$-axis (FIG
\ref{fig:a0potential_negative}).
Therefore there are no vacua in the $\beta < 0$ case.

\subsubsection{$a=1,2$ branches}
\begin{figure}[t]
\begin{center}
\includegraphics[scale=.32]{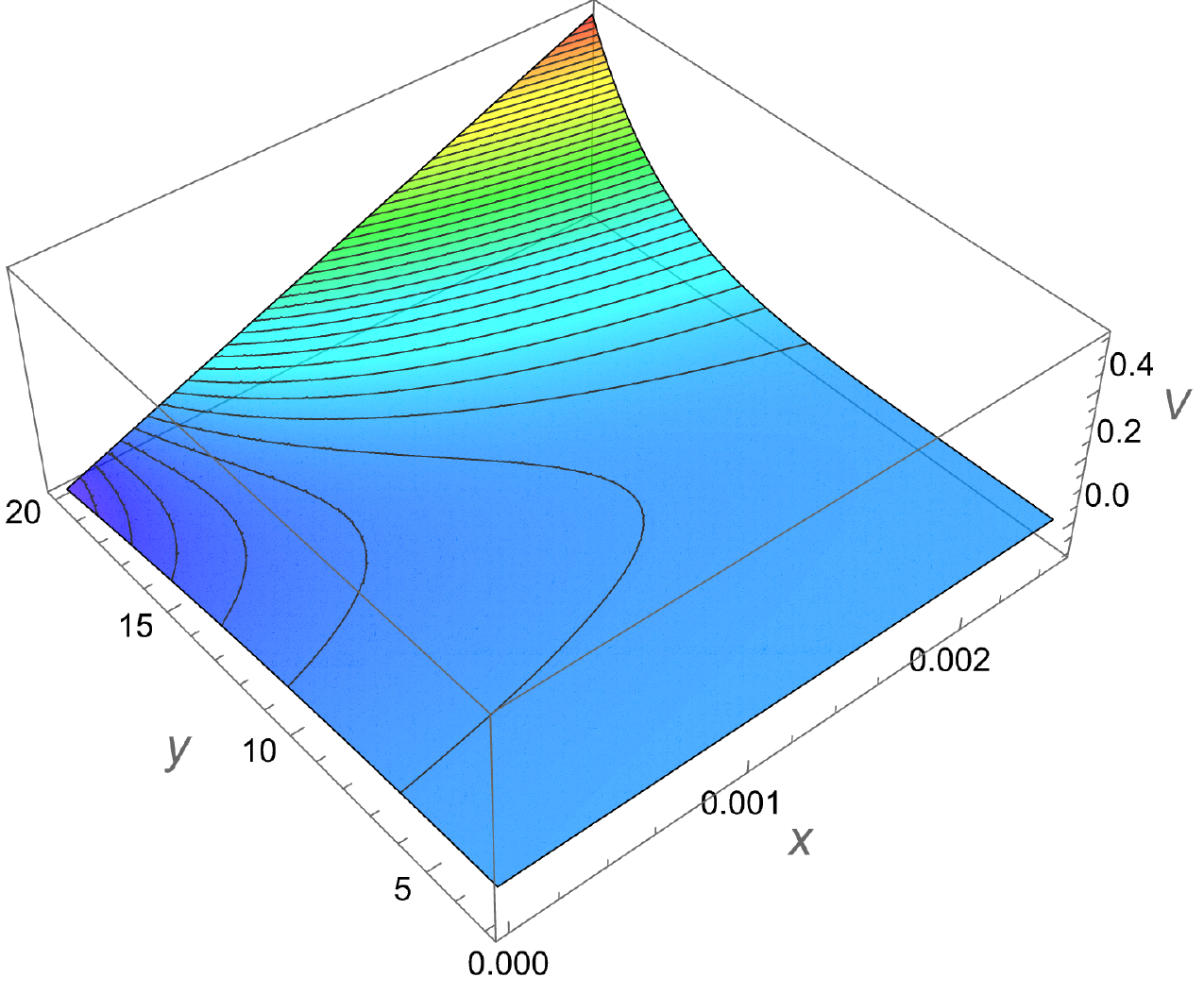}
\includegraphics[scale=.32]{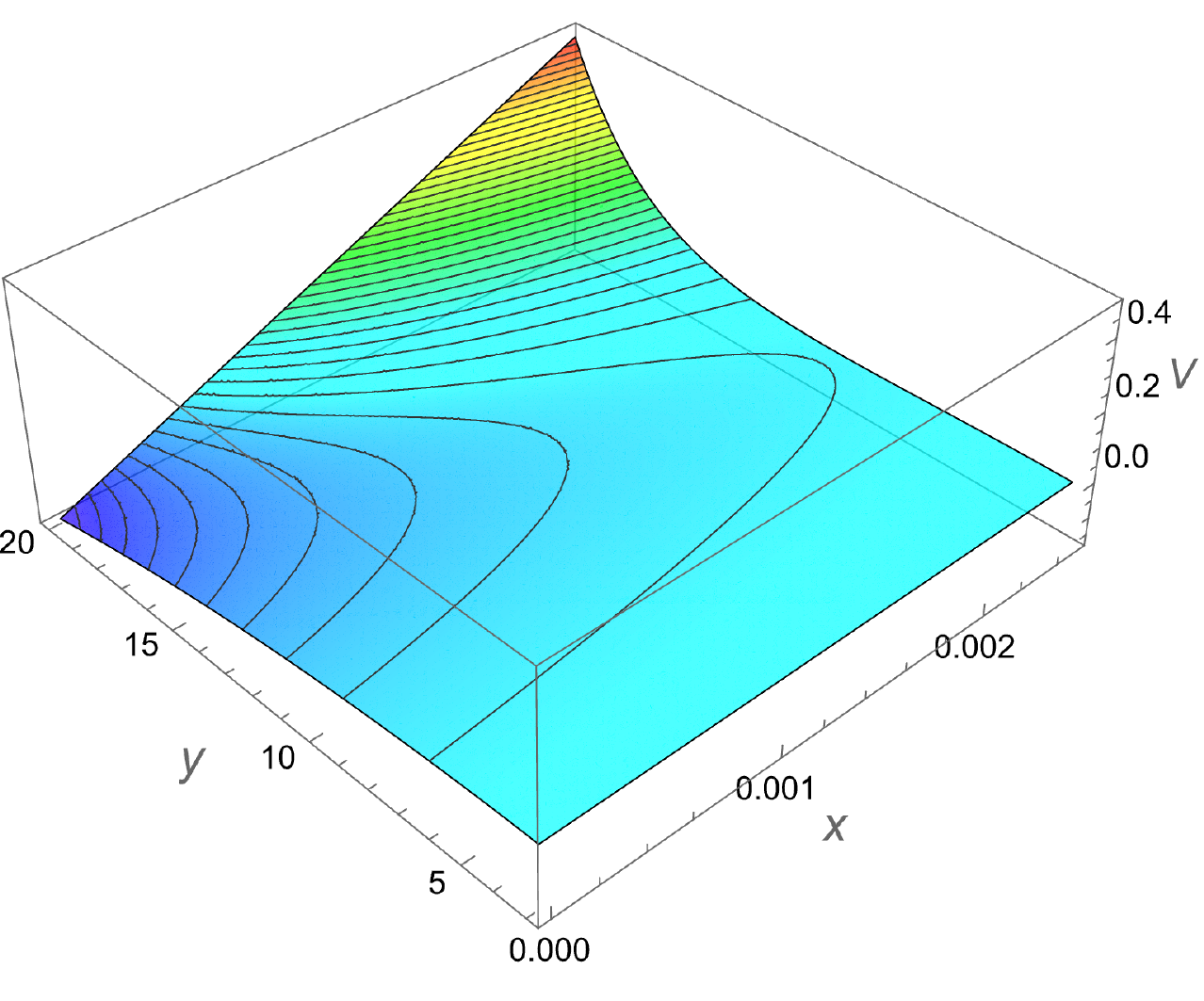}
\end{center}
\caption{The scalar potentials in the $a=1$ (left) and the $a=2$
(right) branches.
The parameter is fixed to $\beta = - 1$.
}
\label{fig:a12potentials_negative}
\end{figure}

For the $a = 1, 2$ branches, 
the condition $D = \left( \frac{q}{2} \right)^2 + \left( \frac{p}{3}
\right)^3 < 0$ is necessary for $F \bar{F}$ to be real numbers.
We find that for $\beta > 0$, this is not the case.
In the case of $\beta < 0$, we find $D <0$ and $F \bar{F} \ge 0$ provided $y \ge 3$.
When this condition is satisfied, the solutions are given by
\begin{align}
F \bar{F} = 2 \sqrt{-\frac{p}{3}}
\cos
\left(
\frac{\theta}{3} + \frac{2 a \pi}{3}
\right)
+
\frac{2}{3} \frac{e^{- \frac{K}{3}}}{32 U}, \quad (a=1,2)
\end{align}
where 
\begin{align}
\tan \theta = - \frac{2 \sqrt{-D}}{q}.
\end{align}
The plots of the scalar potentials are found in 
(FIG. \ref{fig:a12potentials_negative}).
We find that there are infinitely negative directions along $x=0$
both in the $a=1,2$ branches.
Therefore they are unbounded from below and there are no minima in
these branches.

\subsection{Einstein equation}
Given the vacuum energy, we solve the Einstein equation and determine the
spacetime structures.
The equation of motion for the metric is found to be
\begin{align}
&
\Big(
\mathcal{R}_{mn} - \frac{1}{2} g_{mn} \mathcal{R}
\Big)
\notag \\
& \ 
+
\frac{3\alpha + 2\gamma}{24}
\Big(
2 \mathcal{R} \mathcal{R}_{mn} - \frac{1}{2} g_{mn} \mathcal{R}^2 - 2 (\nabla_{m}
 \nabla_{n} - g_{mn} \nabla^2) \mathcal{R}
\Big)
\notag \\
&
+ \frac{3\alpha + 4\gamma}{8}
\Big(
\frac{1}{2} g_{mn} \mathcal{R}_{pq} \mathcal{R}^{pq}
- 2 g^{pq} \mathcal{R}_{m p} \mathcal{R}_{n q}
+ \nabla_{n} \nabla_{p} \mathcal{R}_{m k} g^{k p}
\notag \\
& \ 
- \nabla_{m} \nabla_{p} \mathcal{R}_{k n} g^{k p}
+ \nabla_{k} \nabla_{l} \mathcal{R}_{mn} g^{kl}
+ \nabla_{q} \nabla_{p} \mathcal{R}_{kl} g^{k p}
 g^{l q} g_{mn} \Big)
\notag \\
&
+ 
\frac{\gamma}{4}
\Big(
- \frac{1}{2} g_{mn} \mathcal{R}_{klpq} \mathcal{R}^{klpq} 
+ 2 \mathcal{R}_{m klp} \mathcal{R}_{n} {}^{klp} 
+ 2 \mathcal{R}_{klmp} \mathcal{R}^{kl} {}_{n} {}^{p}
\notag \\
& \ 
+ 4 \nabla^{k} \nabla^{l} \mathcal{R}_{mlnk}
\Big)
- \Lambda g_{mn} = 0,
\label{eq:Einstein_eq}
\end{align}
where $\Lambda$ is the vacuum energy.
Assuming the de Sitter space ansatz;
\begin{align}
\mathcal{R}_{mnpq} = \frac{1}{\lambda^2} (g_{mp} g_{nq} - g_{mq} g_{np}),
\end{align}
where $\lambda$ is the de Sitter radius, 
we have the equation
$\Lambda \lambda^4 - 3  \lambda^2 + 3 \gamma = 0$ from \eqref{eq:Einstein_eq}.
The solutions are found to be
\begin{align}
\lambda^2 = \frac{3  \pm \sqrt{9 - 12  \Lambda \gamma}}{2 \Lambda}.
\label{eq:solutions}
\end{align}
From this expression, we find that there is a solution $\lambda^2 > 0$
when $\gamma>0$ and $\Lambda < 0$, namely, a de Sitter space is
allowed even for the negative vacuum energy $\Lambda < 0$ (the minus
sign in \eqref{eq:solutions}).
Indeed, for $\gamma > 0$ (cf. Eq.~\eqref{eq:component_Lagrangian}), 
a de Sitter space $\lambda^2 > 0$ is allowed in the $\beta > 0$, $a = 0$
branch with the vacuum energy $\Lambda = - 1.1533 \times 10^{-4}$.
Note that the solution \eqref{eq:solutions} is independent of the
coefficient $\alpha$.

\section{Conclusion} \label{sect:conclusion}
In this letter, we studied the four-dimensional $\mathcal{N} = 1$ old
minimal supergravity coupled with the curvature square terms and the 
higher derivative chiral model.
This is a natural model that contains fourth order spacetime
derivatives both in the gravity and the chiral multiplets.

The auxiliary fields play an important role both in the gravity and the
chiral multiplets.
A well-known fact that the auxiliary field $M$ in the gravity multiplet
becomes propagating in the presence of the curvature square terms, and
the non-trivial solutions for $F$ in the higher derivative chiral matter sector, allows us to generate a
non-vanishing scalar potential.
We explicitly showed that the extra scalar field $M$ together with the higher derivative chiral model
generates a scalar potential even in the absence of superpotentials. 
The non-trivial scalar potentials are generated through each solution to the equation of motion of $F$.
Although they are complicated, the explicit forms of the potentials enable
us to find a minimum and the vacuum energy in the $a=0$ branch.
We found that the vacuum generically breaks supersymmetry 
since the auxiliary field $F\bar{F}$ at the vacuum is non-zero in general.
Given the vacuum energy, we solved the Einstein equation
including the curvature squares and found that a (meta) stable de Sitter
space is allowed even for the negative vacuum energy.
This is in contrast to the previous works \cite{Koehn:2012ar, Farakos:2012qu} where no curvature
square terms are present.
Our analysis showed that the de Sitter radius depends on the coefficient $\gamma$ of the Riemann curvature squares.
This would be relevant in the heterotic supergravities \cite{Bergshoeff:1989de, Zwiebach:1985uq} 
where the $\mathcal{R}_{mnkl} \mathcal{R}^{mnkl}$ term plays an important role to cancel the anomaly.

In this letter, we focused on the simplest case, namely, the flat K\"ahler
potential $K = A \bar{A}$ and the constant $U$. 
Even such a case, we found rich structures of scalar potentials. 
This indicates that our findings will be useful for model building without superpotentials.
Since low-energy effective theories generically have non-trivial K\"ahler
potentials and the function $U$, it would be interesting to study vacuum structures
in specific setups like the Skyrme matters \cite{Adam:2011hj,
Adam:2013awa, Gudnason:2015ryh, Gudnason:2016iex}, the D-brane
worldvolume theory \cite{Rocek:1997hi, Sasaki:2012ka}, 
string compactifications \cite{Kachru:2003aw} and so on.
In \cite{Farakos:2012qu}, anti-de Sitter vacua are uplifted to de Sitter
ones by the D-term contributions of gauge sectors.
It is natural to incorporate the derivative corrections to gauge sectors
and study their roles in vacuum structures.
It has been shown that the higher derivative chiral model admits
non-standard supersymmetry breaking vacua such as modulated ground
states 
\cite{Nitta:2017mgk, Nitta:2017yuf, Gudnason:2018bqb, BjarkeGudnason:2018aij}.
It would also be interesting to study the corresponding spacetime
structures in these vacua.

An inflation model based on the higher derivative chiral model has been studied in \cite{Gwyn:2014wna}.
Their analysis shows that there is an intrinsic singularity of the speed of
sound in the $a=2$ on-shell branch in the Einstein supergravity.
It would be interesting to study the inflationary dynamics including the
curvature squared corrections.
We also expect that our analysis may shed light on the complete
understanding of the string landscape and the swampland program
\cite{Susskind:2003kw, Vafa:2005ui}.
We will come back to these issues in future works.

\subsection*{Acknowledgments}
The work of S.S. is supported in part by Grant-in-Aid for Scientific Research (C),
JSPS KAKENHI Grant Number JP20K03952.



\end{document}